\documentclass[usenatbib]{basi}
\usepackage[british]{babel}
\usepackage[varg]{txfonts}
\usepackage{ifpdf}
\ifpdf
  \DeclareGraphicsExtensions{.pdf,.png,.jpg}
  \graphicspath{{./PDF/}}
\else
  \DeclareGraphicsExtensions{.eps}
  \graphicspath{{./EPS/}}
\fi
\begin{document}
\title[CRTS transients]{Discovery, classification, and scientific
  exploration of transient events from the Catalina Real-time Transient Survey}

\author[A.~A.~Mahabal et~al.]%
  {A.~A.~Mahabal$^1$\thanks{email: \texttt{aam@astro.caltech.edu}},
  S.~G.~Djorgovski$^{1,2}$, A.~J.~Drake$^1$, C.~Donalek$^1$,\newauthor
  M.~J.~Graham$^1$, R.~D.~Williams$^1$, Y.~Chen$^1$, B.~Moghaddam$^3$,
  M.~Turmon$^3$,\newauthor
  E.~Beshore$^4$ and S.~Larson$^4$\\
  $^1$California Institute of Technology, 1200 E California Bl.,
      Pasadena, CA 91125, USA\\
  $^2$Distinguished Visiting Professor, King Abdulaziz University, Jeddah,
      Saudi Arabia\\
  $^3$Jet Propulsion Laboratory, Pasadena, CA 91109-8099, USA\\
  $^4$Lunar and Planetary Laboratory, University of Arizona, Tucson, AZ 85721, USA}

\pubyear{2011}
\volume{39}
\setcounter{page}{387}
\pagerange{\pageref{firstpage}--\pageref{lastpage}}
%\pagerange{387--408}

\date{Received 2011 September 12; accepted 2011 October 9}

\maketitle
%------------------------------------------------------------------------------%
\label{firstpage}

\begin{abstract}
Exploration of the time domain -- variable and transient objects and phenomena
-- is rapidly becoming a vibrant research frontier, touching on essentially
every field of astronomy and astrophysics, from the Solar system to cosmology.
Time domain astronomy is being enabled by the advent of the new generation of
synoptic sky surveys that cover large areas on the sky repeatedly, and
generating massive data streams. Their scientific exploration poses many
challenges, driven mainly by the need for a real-time discovery,
classification, and follow-up of the interesting events. Here we describe the
Catalina Real-Time Transient Survey (CRTS), that discovers and publishes
transient events at optical wavelengths in real time, thus benefiting the entire community. We
describe some of the scientific results to date, and then focus on the
challenges of the automated classification and prioritization of transient
events. CRTS represents a scientific and a technological testbed and precursor
for the larger surveys in the future, including the Large Synoptic Survey 
Telescope (LSST) and the Square Kilometer Array (SKA). 
\end{abstract}


%------------------------------------------------------------------------------%
\section{Introduction}\label{s:intro}

Time-domain astronomy is an exciting and rapidly growing research frontier,
ranging from the Solar system to cosmology and extreme relativistic phenomena.
A number of important astrophysical phenomena can be discovered and studied
only in the time domain, e.g.\ supernovae and other types of cosmic explosions.
Variability is observed on time scales ranging from milliseconds to the Hubble
time (by extrapolation). It comes from a broad range of physics, from magnetic
field reconnections to shocks, cosmic explosions, and gravitational collapse.
Time-domain studies often provide important -- or even unique -- insights into
the observed phenomena. There is also a real and exciting possibility of a
discovery of new types of objects and phenomena. Opening new domains of the
observable parameter space often leads to new and unexpected discoveries.

The field has been fueled by the advent of the new generation of digital
synoptic sky surveys, which cover the sky many times, as well as the ability to
respond rapidly to transient events using robotic telescopes. This new growth
area of astrophysics has been enabled by information technology, continuing
evolution from large panoramic digital sky surveys, to panoramic digital
cinematography of the sky. The sky is now a dynamic entity, changing all the
time.

Numerous surveys and experiments have been exploring the time domain at a full
range of wavelengths, and ever more ambitious ones are being planned, most
notably the Large Synoptic Survey Telescope (LSST; \citealt{Ive08}), or the
Square Kilometer Array (SKA) and its precursors. Focusing on the visible
regime, some of the ongoing surveys include, for example, the Robotic Optical
Transient Search Experiment (ROTSE-III; \citealt{Ake03}), the All Sky Automated
Survey (ASAS-3; \citealt{Poj01}), the Palomar Transient Factory (PTF;
\citealt{Rau09}), the Pan-STARRS, \citep{Kai02} and the Skymapper \citep{Kel07},
to name just a few.

Here we describe the Catalina Real-Time Transient Survey, an optical filterless
survey for transients (CRTS; \texttt{http://crts.caltech.edu/}; \citealt{Dra09};
\citealt{Djo11a}). The key motivation behind this project is a systematic
exploration of the time domain in astronomy. CRTS is producing a steady stream
of discoveries, and it also serves as a scientific and technological testbed
for the larger synoptic sky surveys to come.

CRTS is a direct descendant of the Palomar-Quest Event Factory, a real-time
transient detection pipeline that operated as a part of the Palomar-Quest
survey (PQ; \texttt{http://palquest.org/}; \citealt{Djo08}), from 2006~September 
to the end of the survey in 2008~September. Detection of transients, filtering of
artifacts, real-time electronic publishing of events, follow-up strategies,
early efforts on automated classification of events, and many other operational
issues have been developed as a part of that survey, and used as a basis for
the CRTS survey. (We note that the PTF survey also uses essentially the same
operational model, at the same telescope as PQ, but with a much better camera,
and with no real-time publishing of events.)

One key distinguishing feature of the CRTS survey is its open-data policy:
detected transients are published electronically in real time, with no
proprietary period at all, thus enabling a more rapid and diverse follow-up,
and benefiting the entire community. CRTS is perhaps the only major sky survey
so far with such a policy, and we hope to encourage such an approach by other
surveys in the future. As the data rates and volumes continue their
exponential growth, the focus of value shifts to the ownership of expertise,
and not the ownership of the data. Moreover, it is already impossible for any
given group to fully exploit this exponential data richness. The
data-possessive approach is neither efficient nor appropriate.

\begin{figure}
\centerline{\includegraphics[width=12cm]{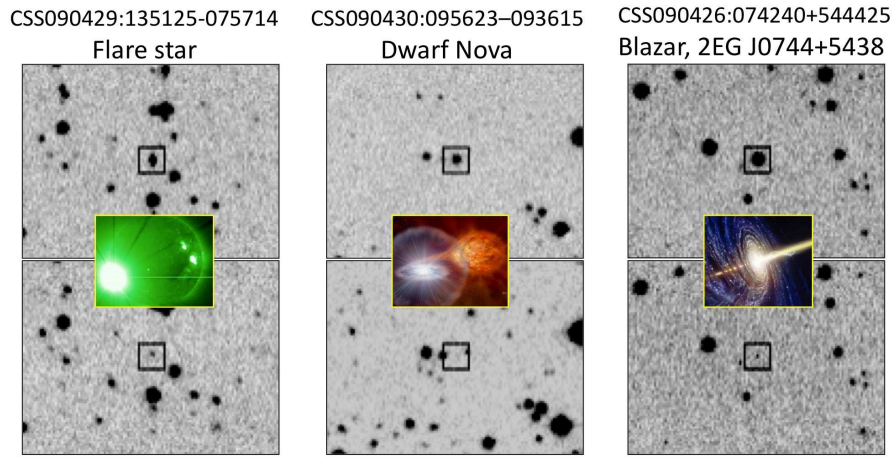}}
\caption{Examples of a few transients from CRTS. Just the discovery images do
not provide enough information for classification. Rapid follow-up is critical
for that purpose. Here, for instance, imaging in multiple filters, spectra and
association with a radio source were used for classification
\citep{Djo11a}.\label{f:crts_transients}}
\end{figure}

In the next few sections we describe briefly the CRTS survey and the process of
detecting transients, and some of the scientific results to date. We then
describe the efforts on automated characterization and classification of these
transients, an important first step for their scientific exploration, and
outline the future possibilities. Fig.~\ref{f:crts_transients} shows a few
examples of transients from CRTS.

%------------------------------------------------------------------------------%
\section{Catalina Sky Survey}\label{s:css}

NASA's Near-Earth Objects Observations Program resulted from a 1998
congressional directive to identify 90\% of near-earth objects (NEOs), which
includes both asteroids and comets $\ge 1$~km in diameter and with a perihelion
distance $<1.3$~AU. This effort is known informally as the Spaceguard goal
\citep{Mor92}. The Catalina Sky Survey (CSS), Mt. Lemmon Survey (MLS), and
Siding Spring Survey (SSS), together referred to as the Catalina Sky Survey
(\citealt{Lar03}; \citealt{Lar06}), has contributed to the Spaceguard mandate by
carrying out a sustained search for NEOs since 2004. Each of Catalina's
three surveys employs telescopes with unique, complementary capabilities, and
are all equipped with identical cameras with 4K$\times$4K, back-illuminated
detectors cooled to cryogenic temperatures. CSS is a 0.68-m f/1.9 classical
Schmidt at Mt. Bigelow, Arizona with a $2.8^{\circ}$ field of view and the
scale of $\sim 2.5''$/pixel, MLS is a 1.5-m f/2 reflector at Mt.
Lemmon, Arizona with a $1.2^{\circ}$ field of view and
$\sim 1.0''$/pixel, and SSS is a 0.5-m f/3 Uppsala Schmidt at
Siding Spring, Australia with a $2.0^{\circ}$ field of view and
$\sim 1.8''$/pixel.

\begin{figure}
\centerline{\fbox{\includegraphics[width=13cm]{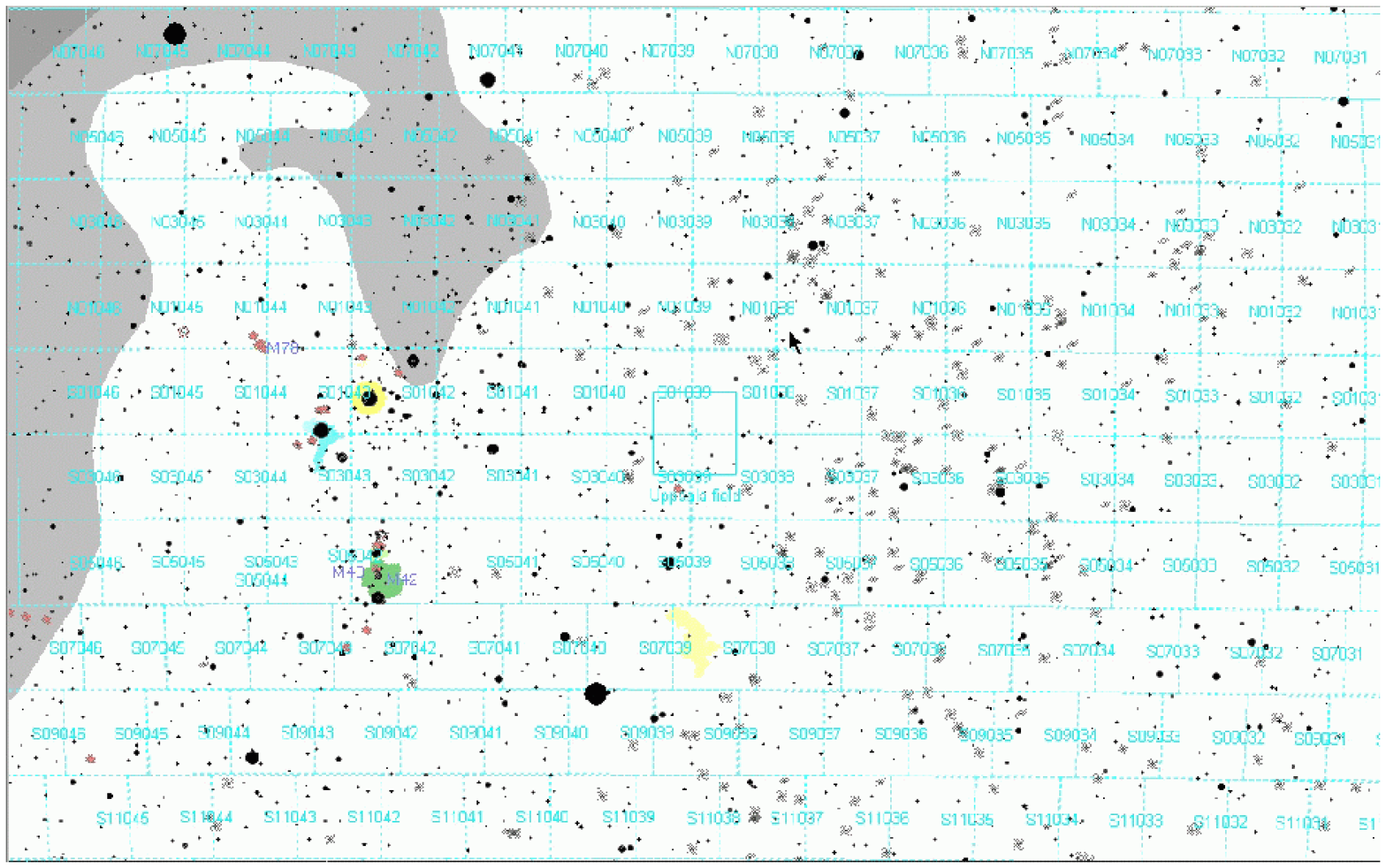}}}
\caption{Catalina obtains images of predefined, standard fields that are unique
to each survey. Here, fields are shown superimposed over
Orion.\label{f:catalina_field}}
\end{figure}

\begin{figure}
\centerline{\fbox{\includegraphics[width=11.2cm]{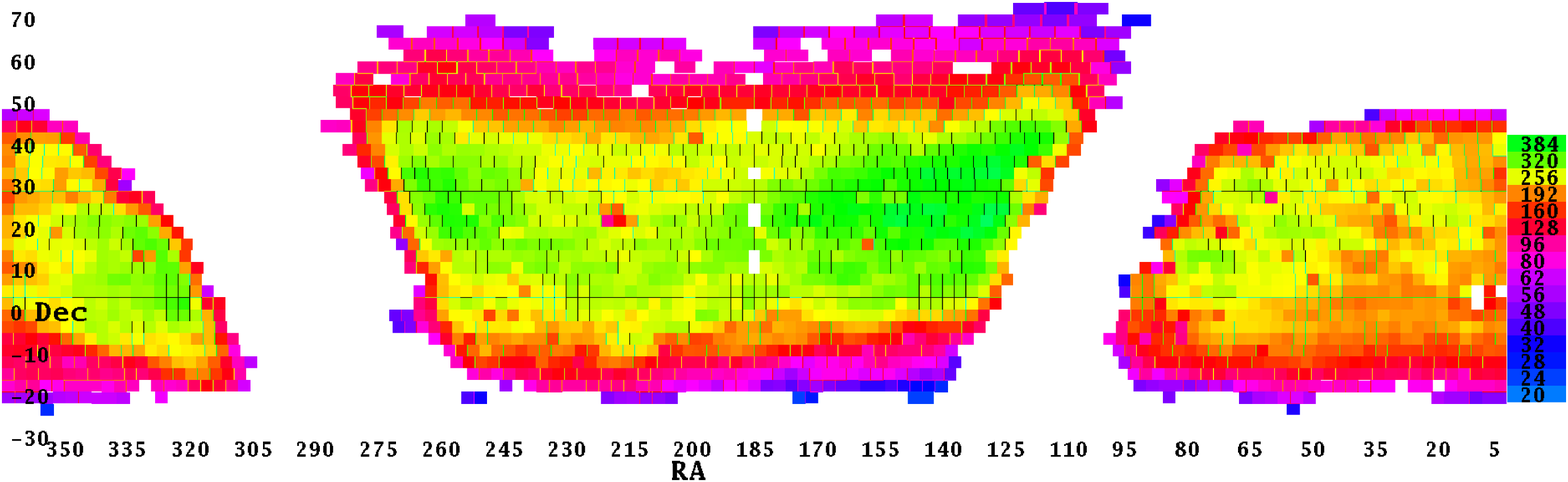}}}
\centerline{\fbox{\includegraphics[width=11.2cm]{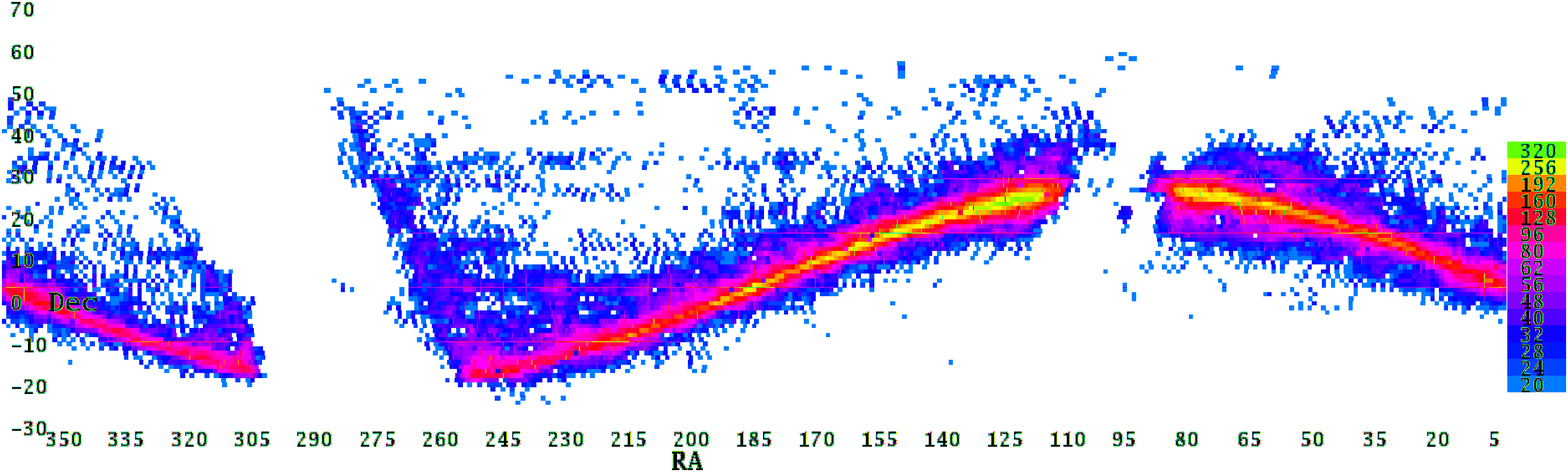}}}
\centerline{\fbox{\includegraphics[width=11.2cm]{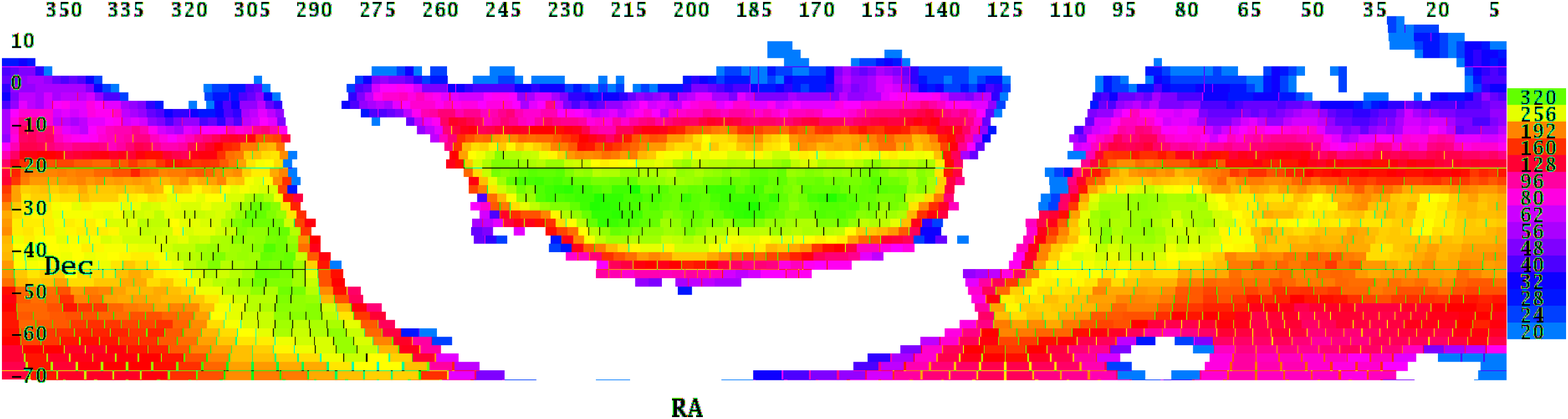}}}
\caption{ Coverage from the 3 CSS telescopes (as of 2011 August). The area
covered and maximum number of epochs for SSS are 15960 sq deg, and 90 nights;
for MLS they are 7238 sq deg. and 81 nights and for CSS the numbers are 24984
sq deg. and 121 nights. With 4 epochs during a night, the maximum number of
epochs for CSS is thus close to 500. Total area for all three surveys with at
least 20 images is 32276 sq. degrees.\label{f:CSSinvAll}}
\end{figure}

The telescopes operate every clear night for about 23 days per lunation.
Predefined, standard fields (see Fig.~\ref{f:catalina_field} for an example)
are observed four times $\sim$10 minutes apart for $\sim$30 seconds with a
small dither between exposures. Observations with CSS are organized to exploit
its medium-faint, wide-field characteristics, and allow complete sky coverage
down to about $-30^{\circ}$ declination in one lunation using 30 second
exposures. SSS often uses a shorter exposure (20 seconds) that allows it to
cover the southern sky south of $-25^{\circ}$ declination each lunation. The
MLS, with a field of view of one square degree, cannot hope to cover the sky
each lunation, and so Catalina exploits its faint-reach, surveying a region
$\pm 10$ degrees along the ecliptic each month using 30--40 second exposures.
All Catalina surveys avoid the Galactic plane, where high star density produces
many false detections and confusing blends ($|b|>10$ for SSS and LMS and
$|b|>20$ for CSS which has a larger plate scale). Statistics compiled by the
NEO Program Office (\texttt{http://neo.jpl.nasa.gov/stats/}) reveals that CSS
has made a significant fraction of all new finds since 2005. Through the most
recently completed half-year of record keeping, CSS has discovered more NEOs
than any other survey and 66 percent of all NEOs discovered since 2005. The
cadence allows us to detect transients varying on timescales from minutes to
years. In addition, the four image sequence provides a significant veto for
asteroids when looking for transients and for artifacts that often cannot be
distinguished from genuine rapid transients in pairs of exposures.
Fig.~\ref{f:CSSinvAll} shows the current sky coverage in the three surveys.
CRTS uses the CSS streams for transient detection.

%------------------------------------------------------------------------------%
\section{Transient detection}

One of the main goals of CRTS has been the detection and characterization of
transients. For our purposes, all genuine non-moving objects that brighten by
a certain amount are transients. These include intrinsic variables (e.g.\
blazars, supernovae) as well extrinsic variable (e.g.\ eclipsing binaries).
Methods and techniques for effective dissemination of alerts were improving in
parallel with the progress of the survey. An important aspect of early
classification is access to additional information about the event either its
past history in the form of images and lightcurves, and/or newer specific
observations. Since follow-up observations are always a bottleneck the
transient detection threshold was kept high initially so that only the blatant
transients will pass through the pipeline.

\begin{figure}
\begin{center}
\begin{tabular}{p{5.5cm}p{7.5cm}}
\centerline{\includegraphics[width=6cm]{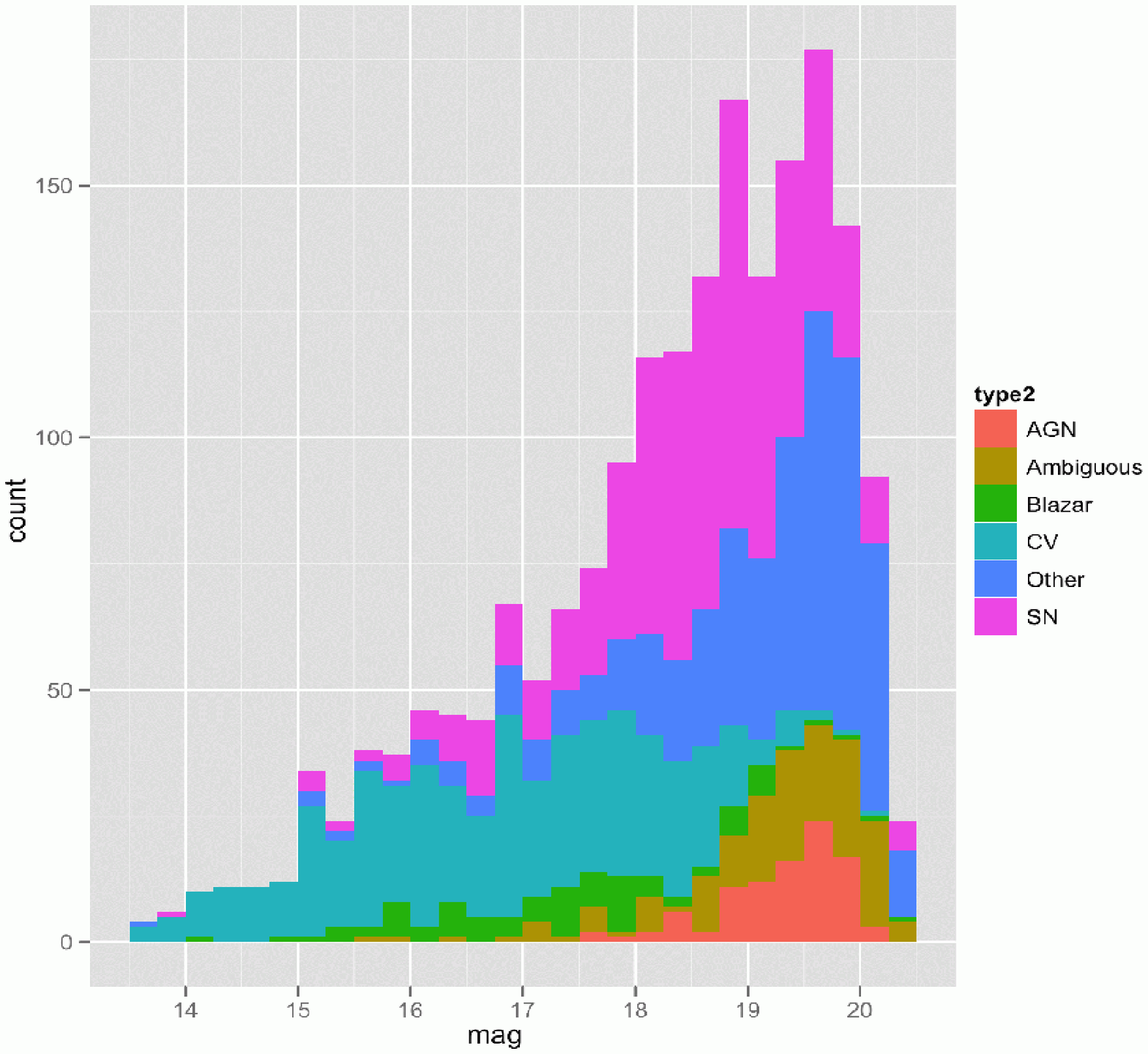}} &
\centerline{\includegraphics[width=6cm]{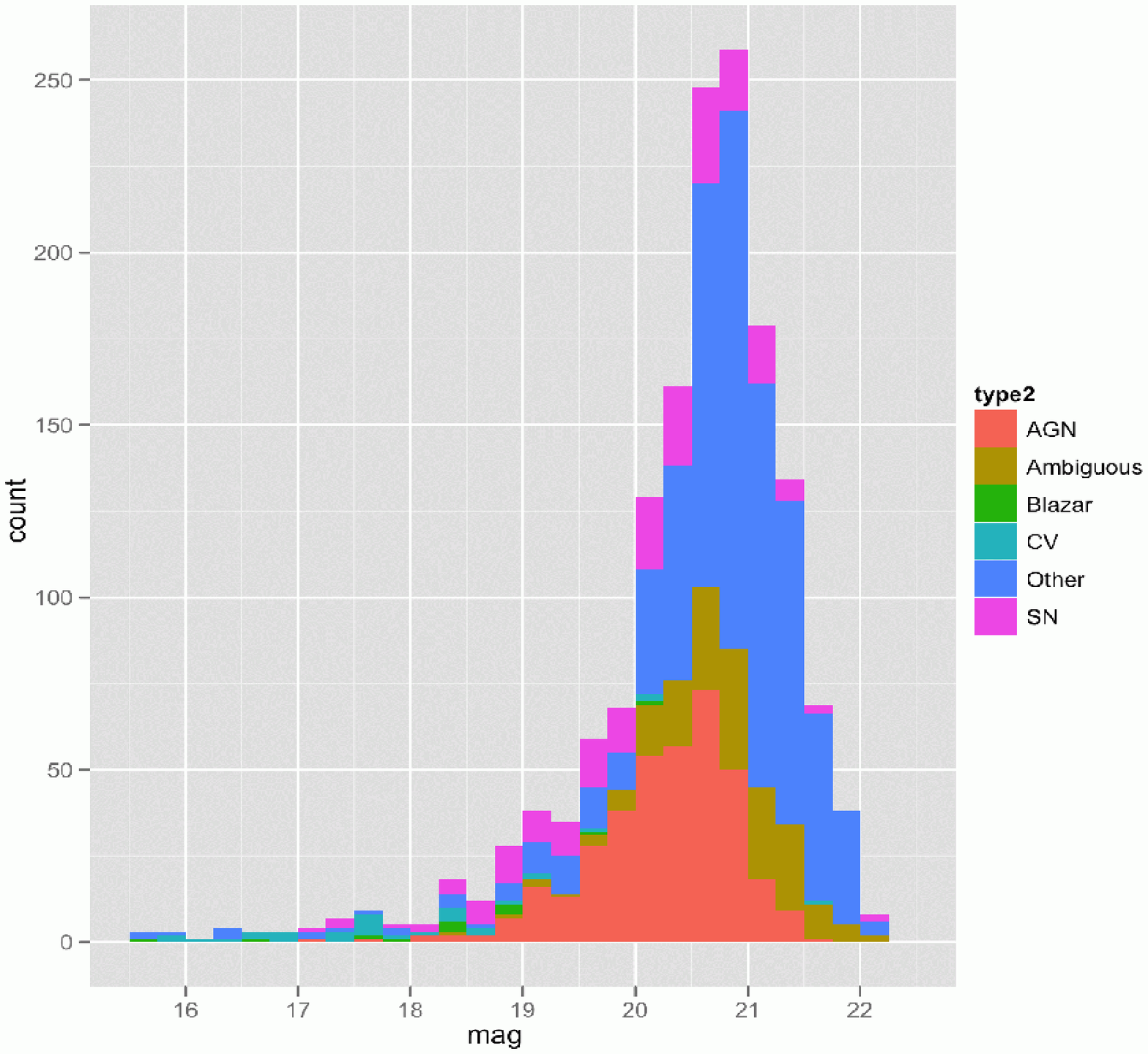}} \\
\centerline{\raisebox{-\height}{\includegraphics[width=6cm]{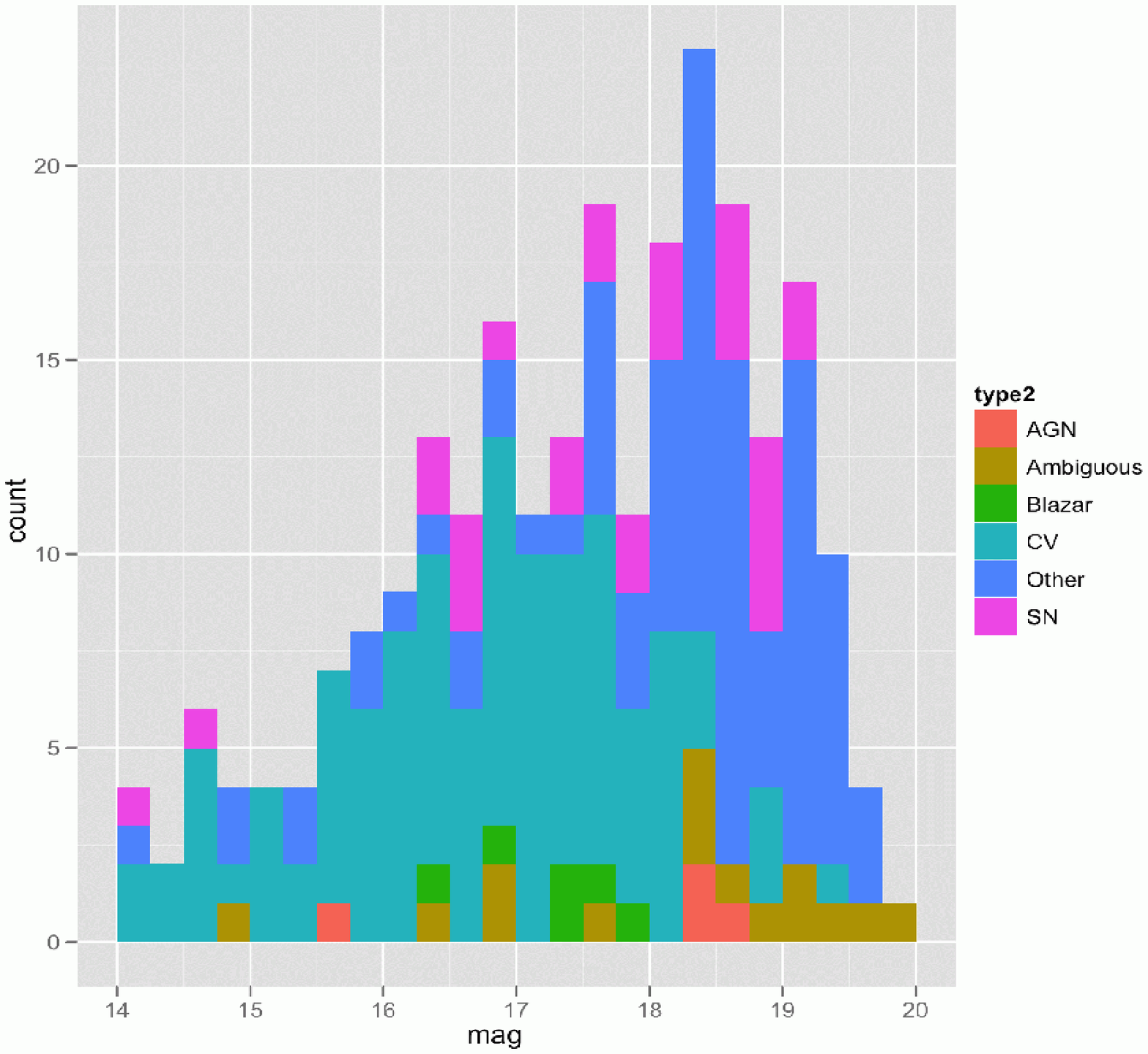}}}
&
\caption{Magnitude distribution for various types of transients found in the
three Catalina surveys. CSS is top left, MLS top right, and SSS bottom left. A
majority of the class labeled as Ambiguous are of type SN/CV i.e.\ when the
brightening of a source makes it cross the transient detection threshold the
historic lightcurve is not unambiguous about possible past brightenings
(something that will rule against a SN), there is no host galaxy (otherwise a
SN is more likely), no nearby radio source (else a blazar is possible). The
distribution of magnitudes of these sources suggest that the SN among them are
typically brighter than average SN and could be associated with dwarf or
fainter galaxies. On the other hand, the CVs in this population would be
fainter than the typical population. \label{f:magdistrib}}
\end{tabular}
\end{center}
\end{figure}

As part of its routine processing CSS uses sextractor to obtain catalogs from
images. Using G-stars in the field the nonfiltered magnitudes are converted to
Johnson $V$. The latest catalogs are compared with corresponding catalogs
obtained for the same area by co-adding at least 20 images from the past. The
deeper co-added image ensures that the comparison is being done with a higher
S/N catalog and thus not many spurious objects and artifacts pass the software
filters. An additional check is done by comparing the catalogs with the higher
resolution catalogs such as from PQ, Sloan Digital Sky Survey (SDSS) and 
the US Naval Observatory (USNO-B). The cadence of taking four
images ten minutes apart is very useful in separating asteroids. Such
asteroids, as well as artifacts, saturations, airplane trails etc. are removed
from potential candidates. After that objects that have brightened
significantly (as much as two magnitudes at the fainter end) are marked as
transients. A cross-check is done with known transients (past outbursts),
radio, X-ray and other catalogs. Typically a few objects per million pass this
threshold. These are published on webpages and alerts sent as VOEvents (see
Sec.~\ref{ss:voevent}) within minutes of the data having been taken. A small
number of artifacts do get through (e.g.\ High Proper Motion (HPM), stars which
are genuine objects but not real transients). We are starting to use an
automated tool to remove these (see Sec.~\ref{s:artifact}), but meanwhile these
are noted after a check by eye and the purer stream posted on a separate
webpage with a lag of few minutes to hours.

\begin{table}
 \centering
  \caption{CRTS Alert statistics as of 2011 August -- some in multiple classes.
  The CV/SN class mentioned here is what forms the bulk of the Ambiguous class
  in Fig.~\ref{f:magdistrib}.}
   \begin{tabular}{@{}ccccccccc@{}}
    \hline
    Tel & All OTs & SNe & CVs & Blazars & Ast/flares & CV/SN & AGN & Other \\
    \hline
    CSS &   2041 & 619 & 507 & 114 & 185 & 274 & 210 & 194 \\
    MLS &   1547 & 193 &  36 &  14 & 124 & 355 & 728 & 217 \\
    SSS &    277 &  28 & 111 &   7 &   5 &  50 &  18 &  60 \\
    Total & 3865 & 840 & 654 & 135 & 314 & 679 & 956 & 471 \\
    \hline
   \end{tabular}
\label{t:disc}
\end{table}

%------------------------------------------------------------------------------%
\section{A sampling of the discoveries}\label{ss:nature}

As shown in Table~\ref{t:disc}, CRTS has been producing various kinds of
transients regularly. These include several types of supernovae (SNe), Cataclysmic
Variables (CV), blazars, Active Galactic Nuclei (AGN), UV Ceti and other
flaring stars, Mira and other high-amplitude variability stars.
Fig.~\ref{f:magdistrib} shows the distribution of some of the more common
classes as a function of magnitude.

An example of a notable CRTS discovery is the type IIn supernovae 2008fz, the
most luminous SN discovered until that time (Fig.~1 of \citealt{Dra10};
\citealt{Gal09}). Another example is the very long-lasting SN 2008iy, a type II
SN, which took over 400 days to reach its peak. Such events possibly originate
in pre-explosion mass loss from the massive $\eta$ Carinae type progenitors
with the SN shock propagating through the stellar wind ejecta for a
considerable time leading to the long rise time.

Another interesting transient is CSS100217:102913+404220 at $z = 0.147$
(\citealt{Dra11b}; Fig.~\ref{f:CSS100217}) with a light curve of a SN IIn, but
making it the most luminous SN ever detected superceding SN 2008fz; the spectra
are consistent with a mix of the pre-explosion Narrow-Line Seyfert 1 (NLS1) AGN, 
and a SN IIn. Hubble Space Telescope (HST) and
Keck AO images reveal that the event occurred within $\sim$150 pc of the
nucleus, well within the narrow-line region. The progenitor could be a massive
star, the formation of which has been long predicted in the 
unstable outer parts of AGN accretion disks \citep{Shl87}; see also
\citet{Jia11}. We are looking in the archival data for more such cases 
of SNe from AGNs.

Since SNe, like all other transients from CRTS are based on change in
magnitudes as ascertained from catalogs, we find more of these that are
associated with faint or dwarf galaxies (see Fig.~\ref{f:dwarf}). These are
likely to represent a population that goes underrepresented in usual
image-subtraction based SN surveys. For more details, see \citet{Djo11b}.

Blazars are often targetted for optical follow-up following their outbursts at
other wavelengths. CRTS provides an unbiased optical monitoring of the entire
sky it covers, and also helps detect new sources. Based on the nature of
variability (Sec.~\ref{s:class}) and association with previously cataloged,
often faint, radio sources we have found several tens of blazar-like sources.
Using the variability of light-curves, we are also searching for counterparts
of unassociated \textit{Fermi} sources \citep{Fer11} by obtaining archival light
curves over several years for all objects in their error ellipses. The data
are being combined with radio data from the Owens Valley Radio Observatory and 
Fermi data. These studies
will provide a better understanding of the radio source population as well as
the types of gamma-ray sources (Mahabal et al., in preparation).

CRTS has discovered more than 500 dwarf nova type CVs, contributing a large
fraction to the known systems. Since many of these are often bright, and the
events get published in real-time, they get regularly followed by small
telescopes (see \citealt{Wil10}, for instance). Similarly, CRTS has discovered
over 100 flare stars (e.g.\ UV Ceti) with some flaring by several magnitudes. It
is important to understand the distribution of these though as a phenomenon
they are fairly well understood. That way the characteristics will allow future
surveys to separate these quickly and go after the rarer phenomena. The flare
stars are easy to catch due to the short cadence of CRTS. Another discovery
this has aided is that of eclipsing white dwarfs where the lightcurve shows a
decrease in brightness as a companion eclipses the white dwarf over a few
minutes. Archival data later revealed several more such systems with low mass
companions \citep{Dra11a}. In addition to these there are a few FU Ori stars
which are seen to continue brightening by several magnitudes over a few years.

We do have an active follow-up program at Palomar, Keck, various telescopes
in India and elsewhere, and we have developed a broad, international network of
collaborations to this end. However, the scientific output of CRTS is
currently limited by the lack of the follow-up, with only a small fraction of
the transients covered (less than 50\% photometrically, and well under 10\%
spectroscopically). This bottleneck (especially in spectroscopy) can only get
worse, as more and larger synoptic surveys come on line.

This brief account is just indicative of the wealth of data produced by CRTS
and the possible resulting projects. Our open-data policy benefits the entire
astronomical community, generating science now, and preparing us for the larger
surveys to come.

%------------------------------------------------------------------------------%
\section{Characterization and classification techniques}\label{s:class}

\begin{figure}
\centerline{\includegraphics[width=13cm]{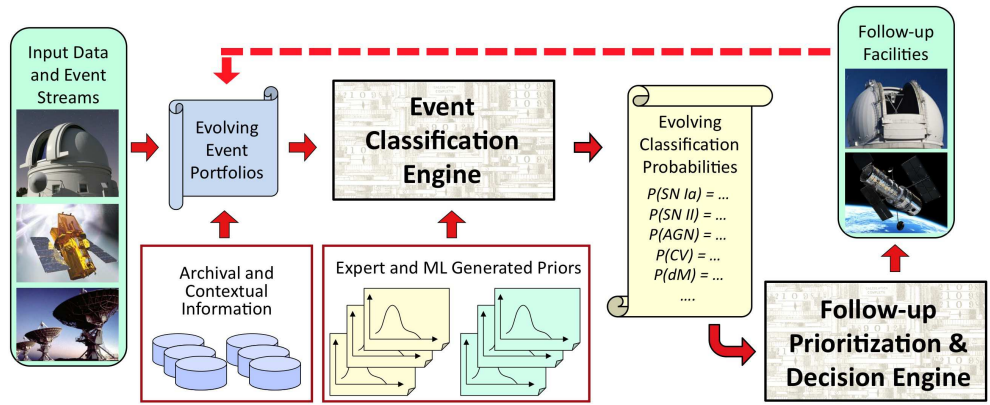}}
\caption{ An overall conceptual outline of the classification system including
transient detection, dissemination, and feedback. The initial input consists of
the generally sparse data describing transient events discovered in sky surveys
(e.g.\ magnitudes and sky positions). These are supplemented by archival
measurements from external, multi-wavelength archives corresponding to this
spatial location, if available (e.g.\ radio flux and distance to nearest
galaxy). Both are collected in evolving electronic portfolios containing all
currently available information for a given event. These data are fed into the
Event Classification Engine; another input into the classification process is a
library of priors giving probabilities for observing these particular
parameters if the event belongs to a class $y$. The output of the
classification engine is a set of probabilities of the given event belonging to
various classes of interest, which are updated as more data come in, and
classifications change. This forms an input into the Follow-up Prioritization
and Decision Engine. It would prioritize the most valuable follow-up
measurements given a set of available follow-up assets (e.g.\ time on large
telescopes, Target-of-Opportunity observations, etc.), and their relative cost
functions. What is being optimized is: (a) which new measurements would have a
maximum discrimination for ambiguous classifications, and/or (b) which
follow-up measurements would likely yield most interesting science, given the
current best-guess event classification? New measurements from such follow-up
observations are fed back into the event portfolios, leading to dynamically
updated/iterated classifications, repeating the cycle. \label{f:overall} }
\end{figure}

To understand the classification of transients, it is instructive and necessary
to look at the bigger picture involving other modules. Fig.~\ref{f:overall}
shows a schematic which places classification in the centre and interacting
with original observations, prior information, feedback etc. We will look at
all these in turn.

The usual scientific measurement and discovery process  operates on time scales
from days to decades after the original measurements, feeding back to a new
theoretical understanding. However, that clearly would not work in the case of
phenomena where a rapid change occurs on time scales shorter than what it takes
to set up the new round of measurements. This results in the need for
real-time systems, consisting of computational analysis and decision engine,
and optimized follow-up instruments that can be deployed selectively in (or in
near) real-time, where measurements feed back into the analysis immediately.
The requirement to perform the analysis rapidly and objectively, coupled with
massive and persistent data streams, implies a need for automated
classification and decision making. VOEvents are used for dissemination of
transient events and as the transport between the different components of the
classification system.

\begin{figure}
\centerline{\includegraphics[width=6.5cm]{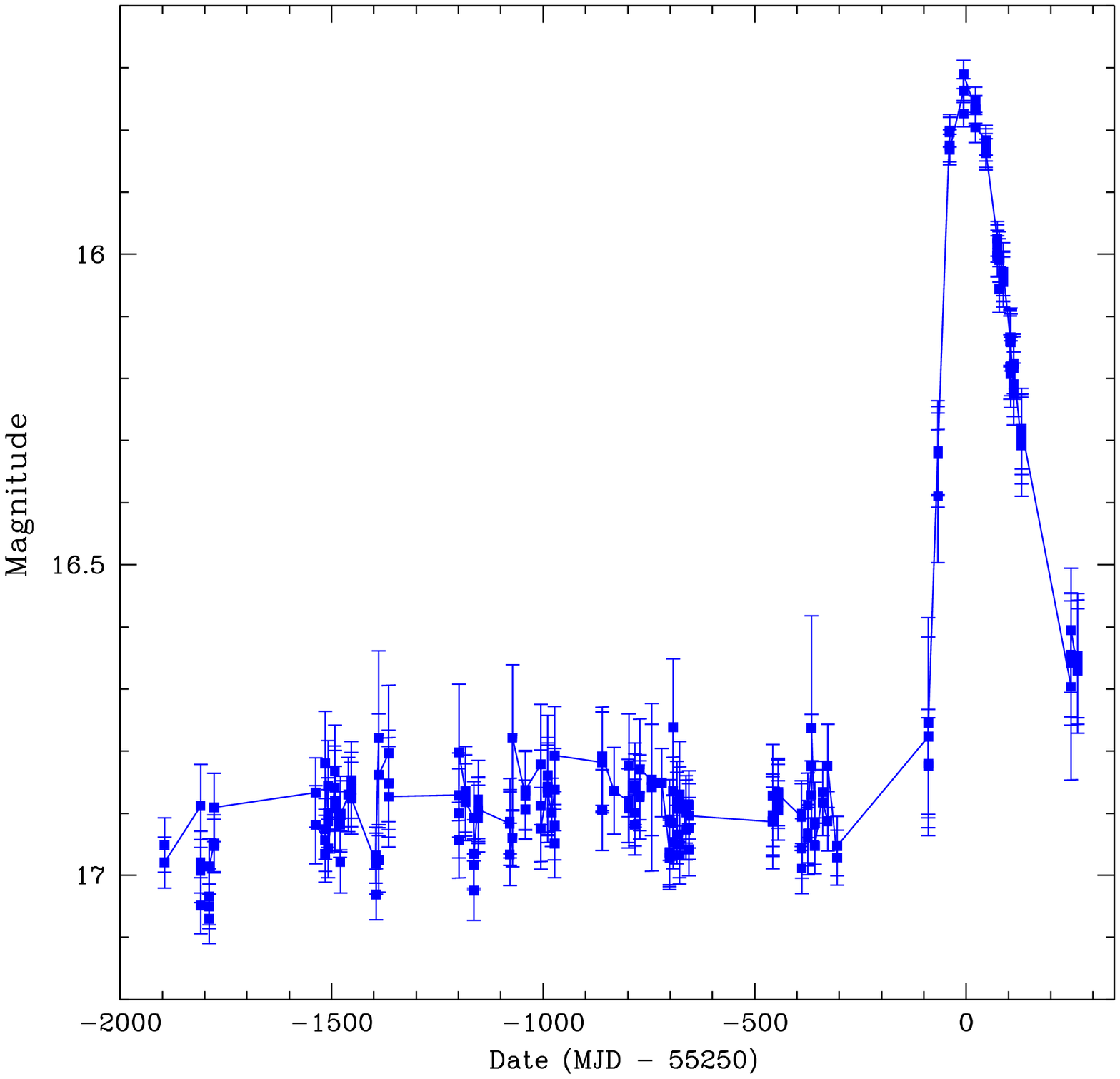} \qquad
\includegraphics[width=6.5cm]{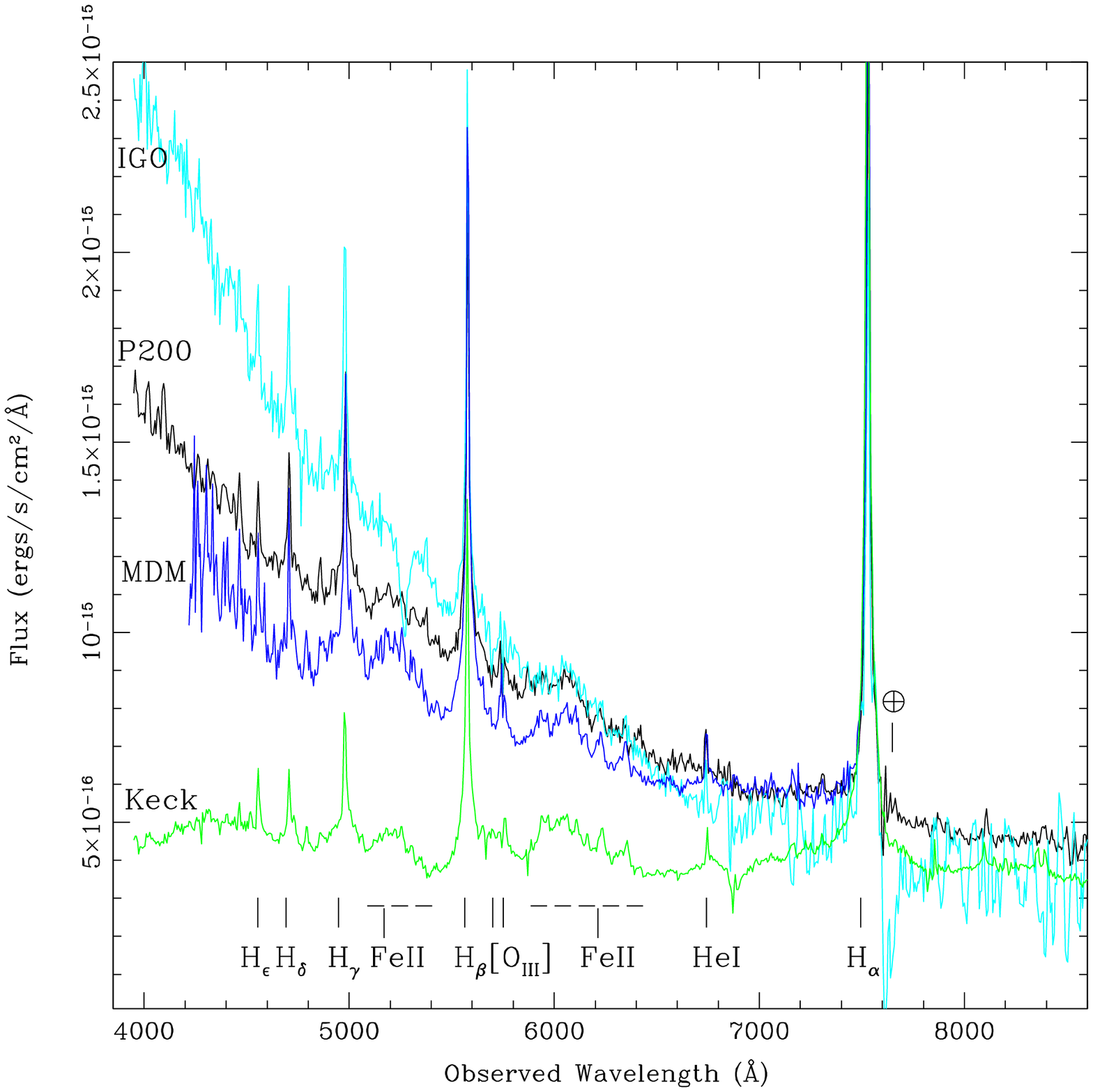}}
\caption{ The remarkable transient CSS100217:102913+404220, the most luminous
Supernova (type IIn) known to date, associated with an AGN galaxy. This may be
the first example of long-predicted supernovae associated with the unstable
outer regions of AGN accretion disks \citep{Dra11b}. Left: the CRTS light
curve; right: evolving spectra of the outburst, showing a combination of the
narrow-line Seyfert 1 (as observed by SDSS, pre-explosion) and a Type IIn SN.}
\label{f:CSS100217}
\end{figure}

The broad classification mantra involves: (1) for the given transient obtain
contextual information, (2) using that and the discovery parameters, determine
probabilities of it belonging to various classes using priors, (3) obtain
follow-up to best disambiguate competing classes, (4) feedback the observations
and repeat until reaching a threshold probability or determining it to be a
less than interesting transient.

In this section we describe the various classification techniques based on a
variety of parameters including contextual information; the use of citizen
science; a fusion module to combine the confidences of the different
classifiers objectively, and the event publication mechanism.

\begin{figure}
\centerline{\includegraphics[width=13cm]{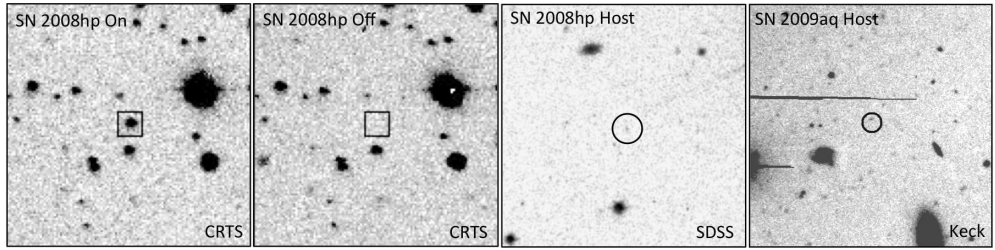}}
\caption{Examples of the extreme dwarf galaxy hosts of luminous SNe. The first
two panels show the images of SN 2008hp = CSS081122:094326+251022 at the
discovery epoch, and after it has faded away. The next panel shows a zoom-in on
the SDSS image of the field; the $\sim 23$ mag host galaxy is circled,
corresponding to the absolute magnitude $M_r \approx -12.7$ mag. The last panel
shows the confirmed $\sim 23$ mag host galaxy (circled) of SN 2009aq =
CSS090213:030920+160505, with the absolute magnitude $M_r \approx -13$ mag.
Measurements of star formation rates and metallicities in these extreme dwarf
hosts will help us understand their extreme specific SN rates, and the
propensity to host ultra-luminous SNe. } \label{f:dwarf}
\end{figure}

\subsection{Artifact removal}\label{s:artifact}

\begin{figure}
\centerline{\includegraphics[width=13cm]{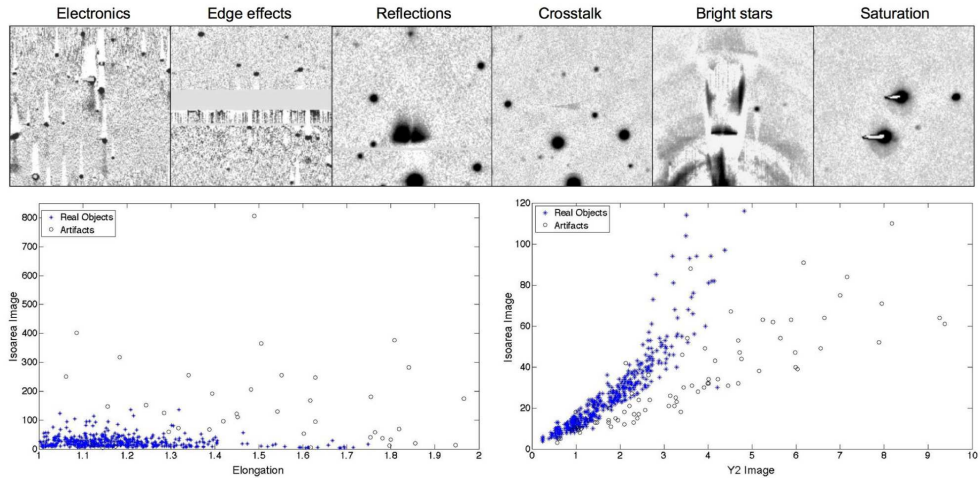}}
\caption{Automated classification of candidate events for PQ data, separating
real astronomical sources from a variety of spurious candidates (instrument
artifacts). Image cutouts on the top show a variety of instrumental and data
artifacts which appear as spurious transients, since they are not present in
the baseline comparison images. The two panels on the bottom show a couple of
morphological parameter space projections, in which artifacts (circles)
separate well from genuine objects (asterisks). A multi-layer perceptron (MLP)
ANN is trained to separate them, using 4 image parameters, with an average
accuracy of $\sim 95\%$. See \citet{Don08} for more details. \label{f:artifact}}
\end{figure}

A first step in classification is to separate genuine objects from artifacts.
We have successfully demonstrated such separation with the PQ Survey data. The
base of knowledge is built by experts looking at a subset of the images and
visually classifying the objects as `real' or `artifact'. Such a dataset is
then used to train a supervised machine learning algorithm (e.g.\ a Neural
Network and/or a Decision Tree) in order to have an automatic classification
that allows us to reject the false positives that the pipeline passes as
transients (see Fig.~\ref{f:artifact}). More details can be found in
\citet{Don08}. We will be implementing artifact classification with CRTS data.

\subsection{Bayesian event classifier}\label{ss:BN}

\begin{figure}
\centerline{\includegraphics[width=13cm]{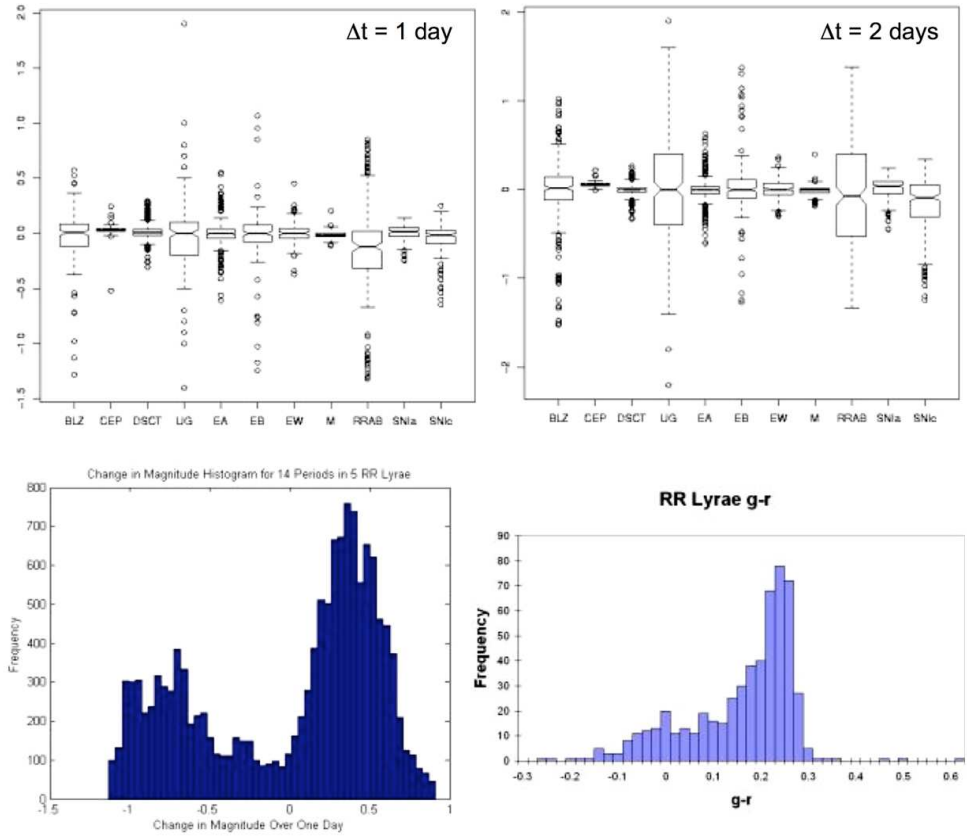}}
\caption{ Examples of prior distributions of selected observables for different
types of astrophysical variable sources compiled from the literature, and
processed by us. Top: box plots of flux variability amplitudes for different
types of objects (plotted along the X axis), sampled with time baselines of 1
day (left) and 2 days (right). There are clear qualitative differences in
behavior among different types of objects, and they depend on time baseline.
The bottom row shows the prior distributions for one particular type of
variable sources, the RR Lyrae stars, with flux (magnitude) change after one
day (left), and colour (right). \label{f:prior} }
\end{figure}

The main astronomical inputs available for classification are in the form of
observational and archival parameters for individual objects, which can be put
into various, often independent subsets. Examples of parameters include various
fluxes at different wavelength or wavelength bands, associated colours or
hardness ratios, proximity values, shape measurements, magnitude
characterizations at different timescales. The heterogeneity and sparsity
of data  make the use of Bayesian methods for classification a natural choice.
Distributions of such parameters need to be estimated for each type of variable
astrophysical phenomenon that we want to classify (Fig.~\ref{f:prior}). This
knowledge is bound to be incomplete and will have to be gradually updated.
Then an estimated probability of a new event belonging to any given class can
be evaluated from all of such pieces of information available, as described
below. Let us denote the feature vector of event parameters as $x$, and the
object class that gave rise to this vector as $y$, $1 \le y \le K$, where $K$
is the total number of classes. While certain fields within $x$ will almost
certainly be known, such as sky position and brightness in selected filters,
many other parameters will be known only selectively: brightness change over
various time baselines, and object shape.

\begin{figure}
\centerline{\includegraphics[width=8.5cm]{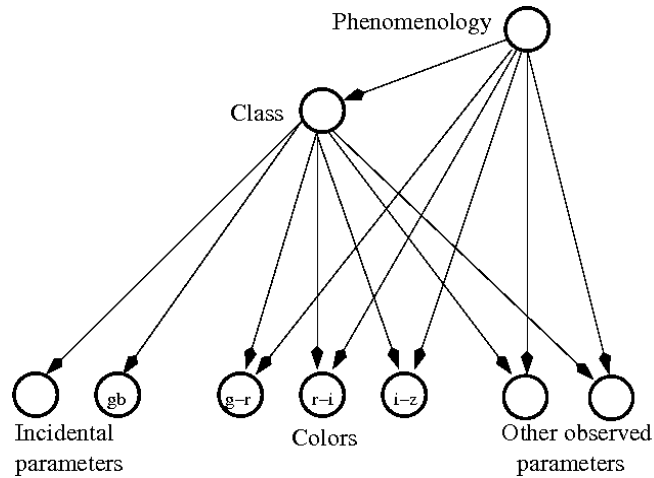}}
\caption{ A schematic diagram of the preliminary BN based on colours and
contextual information as described in the text. There are only 6 classes of
variable objects considered, one of which is composite of any objects not
captured in the first 5, thus serving as a model for hypothetical previously
unknown types. The `Phenomenology' to `observed parameters' connections
indicate possible inputs from theory. The actual BN implementation proposed
here would have many more classes of objects and many more types of observable
parameters. The basic classification nevertheless provides another check for
selecting the best candidates for spectroscopy. We are working on combining
this with another Bayesian tool based on lightcurve data for more accurate
classification. \label{f:BN}}
\end{figure}

The parameters can be divided into several subsets based on similarity and
interdependence. This decoupling is advantageous in two ways. First, it
allows us to circumvent the `curse of dimensionality,' because we will
eventually have to learn the conditional distributions $P(x_b | y = k)$ for
each $k$. As more components are added to $x_b$, more examples will be needed
to learn the corresponding distribution. The decomposition keeps the
dimensionality of each block manageable. Second, such decomposition allows us
to cope easily with ignorance of missing variables. We simply drop the
corresponding sets. As a simple demonstration of the technique, we have been
experimenting with a prototype Bayesian Network (BN) model, schematically
illustrated in Fig.~\ref{f:BN}. See \citet{Mah08} for more details.

We use a small but homogeneous data set involving colours of transients detected
in the CRTS survey, as measured at the Palomar 1.5-m telescope (hereinafter
referred to as P1.5m). We have used
multinomial nodes (discrete bins) for 3 colours, with provision for missing
values, and a multinomial node for Galactic latitude which is always present
and is a probabilistic indicator of whether an object is Galactic or not. The
current priors used are for five distinct classes: cataclysmic variables (CVs),
supernovae (SN), Blazars, other AGNs, UV Ceti stars and all else bundled into a
sixth class, called Rest.
%
%This toy model produces a completeness of over 70\% for the two main classes.
%
Using a sample of 316 SNe, 277 CVs, and 104 blazars, and a single epoch
measurement of colours, in the relative classification of CVs vs. SNe, we obtain
a completeness of $\sim 80\%$ and a contamination of $\sim 19\%$, which reflects
a qualitative colour difference between these two types of transients. In the
relative classification of CVs vs. blazars, we obtain a completeness of
$\sim 70{-}90\%$ and a contamination of $\sim 10{-}24\%$ (the ranges corresponding to
different BN experiments), which reflects the fact that colours of these two
types of transients tend to be similar, and that some additional discriminative
parameter is needed. These numbers are based on a single epoch (up to four
bands besides the incidental parameters) and will improve further as the priors
improve. Eventually we will use a BN with an order of magnitude more classes,
more parameters, and additional layers. The end result will be the posteriors
for the {\it Class} node from the marginalized probabilities of all available
inputs for a given object.

Prior distributions of various observable parameters -- like those used in the
BN described above -- are being put together for a variety of distinct
astrophysical variable sources using the initial event measurements from the
survey pipeline, corresponding data from the federated VO archives, and our own
measurements obtained in the CRTS survey and its follow-up observations. The
parameters for which we are building (and subsequently, updating) priors
include primarily colours, light curves (flux histories) sampled at different
time baselines (e.g.\ measurements separated by an hour, from night to night,
etc.), r.m.s.\ and maximum flux variations  etc., conditional on object type
such as type Ia Supernova. The priors come from a set of observed parameters
like distribution of colours, distribution of objects as a function of Galactic
latitude, frequencies of different types of objects etc. The posteriors we are
interested in include determining the type of an object based on, say, its ($r{-}i$)
colour, Galactic latitude and proximity to another object.

\subsection{Light curve classification}\label{ss:dmdt}

\begin{figure}
%\begin{center}
%
%\begin{tabular}{p{6.3cm}p{6.3cm}}
%
\includegraphics[width=6.3cm]{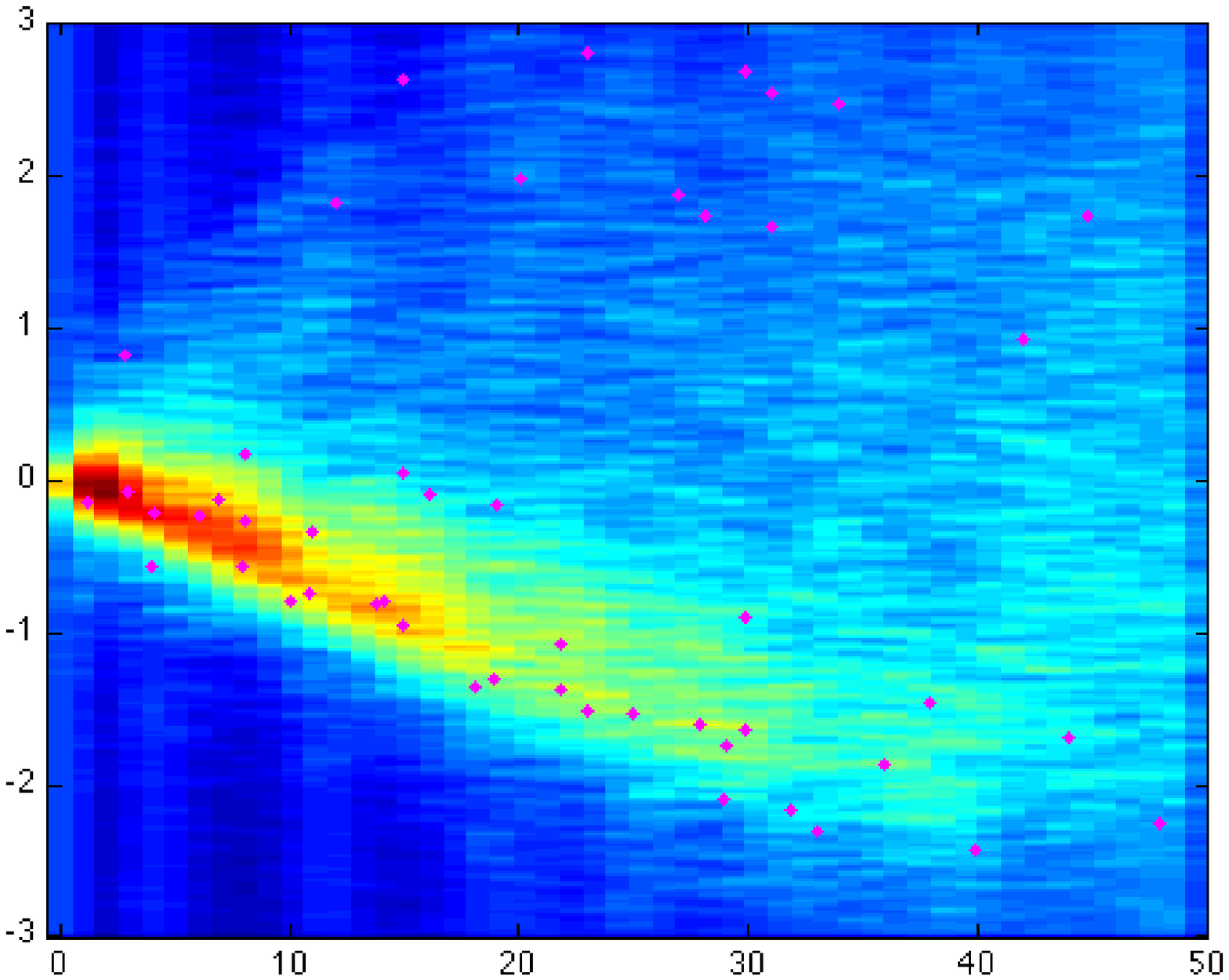} \qquad
\includegraphics[width=6.3cm]{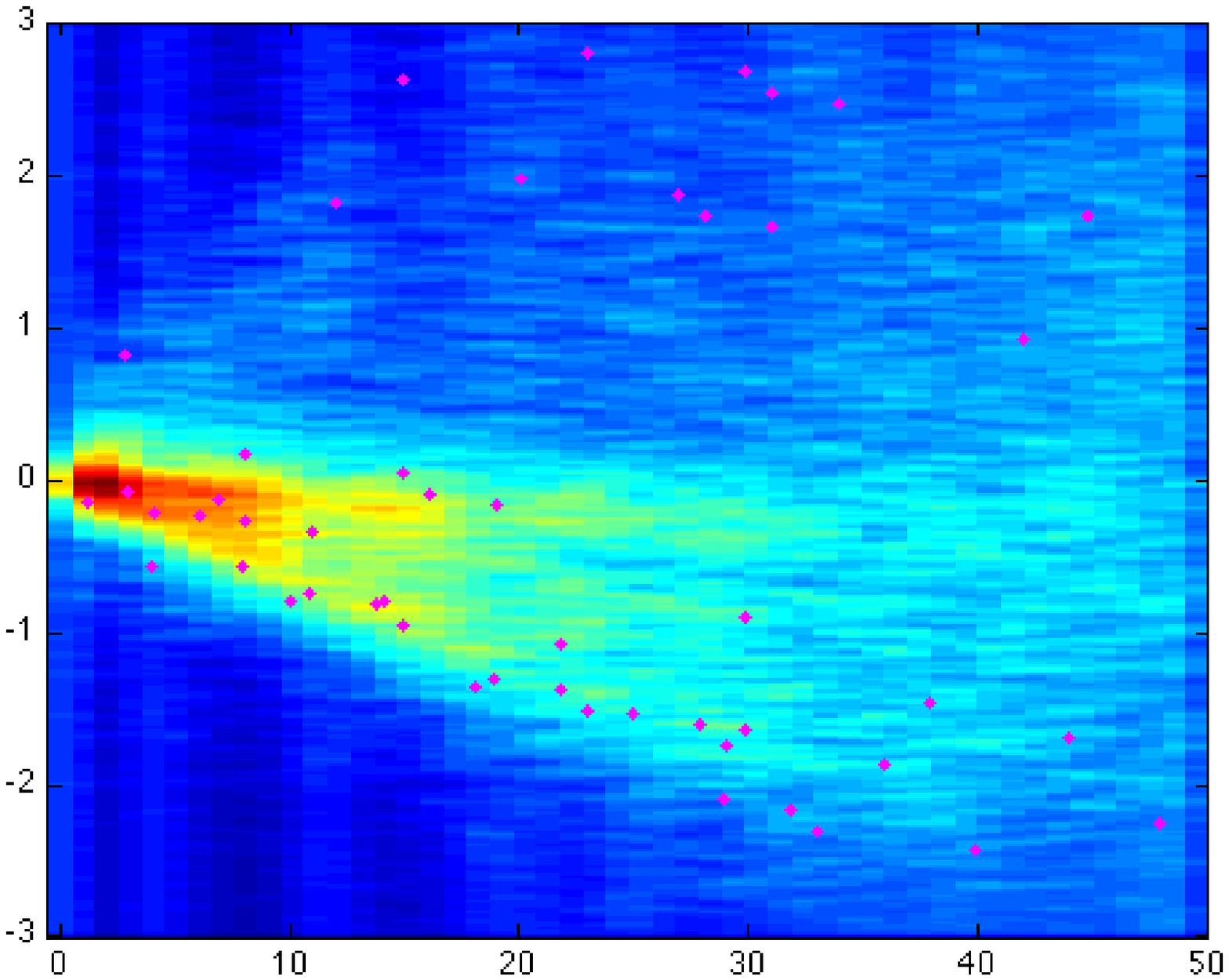} \\
%\raisebox{-\height}{
\includegraphics[width=6.3cm]{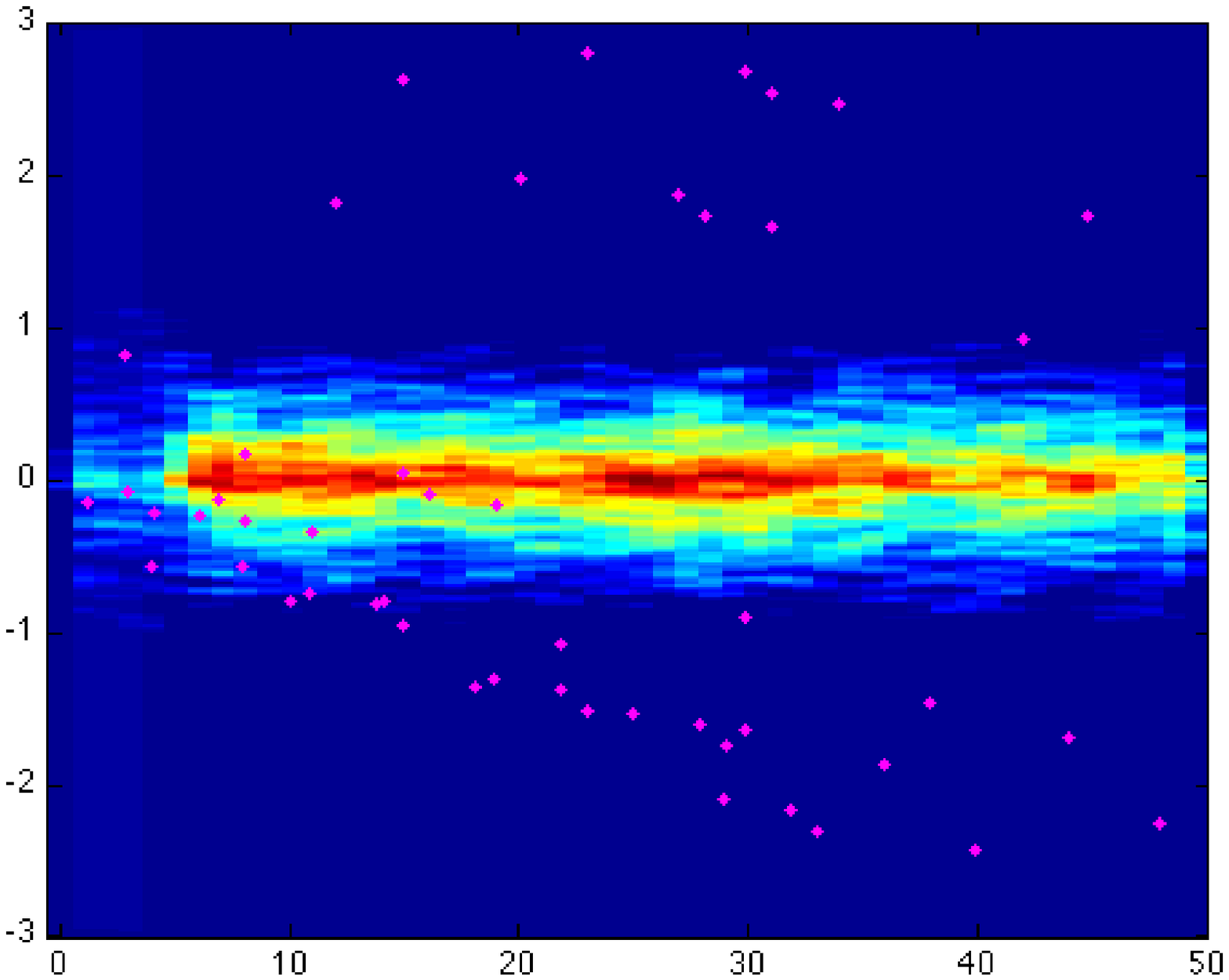} \qquad
\includegraphics[width=5.8cm]{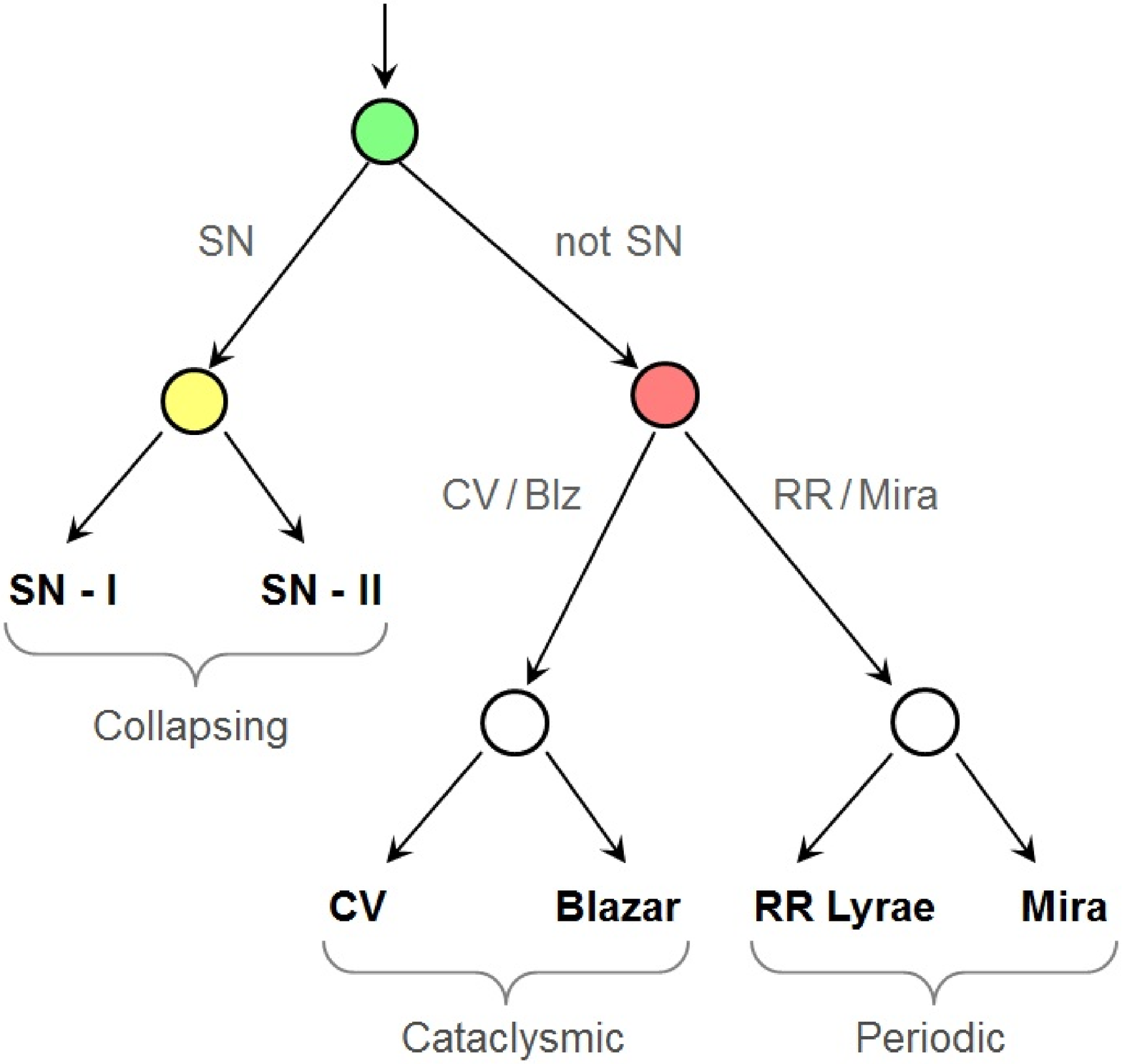}
%} &
%
\caption{ Examples of ($dm, dt$) Probability Distribution Functions. Smoothed
2D histograms are shown for SN Ia (top-left), SN IIP (top-right) and RR Lyrae
(bottom-left), using bins of width $\delta t = 1$ day ($x$-axis), and
$\delta m = 0.01$ ($y$-axis).
%
%The histograms were smoothed with a 3-tap triangular $\delta t$ kernel $= [0.25
%0.5 0.25]$ and a Gaussian $\delta m$ kernel of FWHM $= 0.05$ mag.
%
The superimposed diamonds are from a single LC (of SN Ia). PDFs for the two SN
types form a better fit than that of RR Lyrae (and SN Ia is a better fit than
SN II P). Various metrics on probability distributions can be used to
automatically quantify the degree of fitness. The decision tree used is shown
at bottom-right. \label{f:dmdt}}
%
%\end{tabular} \end{center}
\end{figure}

When it comes to sparse and/or irregular light curves (LC) for any given object
class the structure may not be obvious to the eye. However the salient
features can be exploited by automated classification algorithms. In
particular, by pooling LCs for different objects belonging to a class we can
effectively represent and encode this characteristic structure
probabilistically in the form of an empirical probability distribution function
(PDF) that can be used for subsequent classification of a LC with even a few
epochs. Moreover, this comparison can be made incrementally over time as new
observations become available, with our final classification scores growing
more confident with each additional set of observations. This forms the basis
for a real time classification methodology. Since the observations come in the
form of flux at a given epoch, for each point after the very first one we can
form a ($\delta m, \delta t$) pair. We focus on modeling the joint distribution
of all such pairs of data points for a given LC. By virtue of being
increments, the empirical probability density functions of these pairs are
invariant to absolute magnitude and time shifts, which is desirable in building
a stable feature representation of LCs for classification algorithms to use.
Additionally, these densities conveniently allow upper limits to be encoded as
well, e.g.\ forced photometry magnitudes at a supernova location in images
taken before the star exploded. We currently use smoothed 2D histograms to
model the distribution of elementary ($dm, dt$) sets. This is a computationally
simple yet effective way to implement a non-parametric density model that is
flexible enough for object classes. Fig.~\ref{f:dmdt} shows the joint 2D
histograms for 3 classes of objects and how a given candidate LC measurements
fit these 3 class-specific histograms. In our preliminary experimental
evaluations with a small number of object classes (single outburst like SN,
periodic variable stars like RR Lyrae and Miras, as well as stochastic
variables like blazars and CVs) we have been able to show that the density
models for these classes are potentially a powerful method for object
classification from sparse/irregular time series as typified by observational
LC data.

Currently we are using the ($dm, dt$) distributions for classification in a
binary mode i.e.\ successive two-class classifiers in a tree structure (see
bottom-right part of Fig.~\ref{f:dmdt}). SNe are first separated from non-SNe
(the easiest bit, currently performing at 98\%), then non-SNe are separated
into stochastic versus non-stochastic, and then each group further separated
into more branches. The most difficult so far has been the CV-blazar node
(based on just the ($dm, dt$) density i.e.\ without bringing in the proximity to
a radio source since we are also interested in discovering blazars that were
not active when the archival radio surveys were done). Currently it is performing
at 71\%. We are also exploring Genetic Algorithms to determine the optimal dm
and dt bins for different classes. This will in turn advise follow-up observing
intervals for specific classes.
%
%In a Bayesian approach, x and y are related via
%
%$$ P(y=k|x) = P(x|y=k)P(k)/P(x)    \\ \alpha P(k)Px|y=k) \\ \approx
%P(k)\Pi_{b=1}^B P(x_b|y=k) $$

\subsection{Follow-up}

There are several reasons why follow-up observations for the transient
candidates are crucial. (1)~Since CRTS does not employ filters, no colour
information is available for the transients when they are first detected. Since
colours are often necessary to distinguish between different classes, we need to
obtain these from elsewhere. (2)~Since the purpose of the CSS survey is looking
for asteroids, we cannot rely on it for repeat observations at specific times
that we may need them. One of the expected outcomes of the ($dm, dt$)
classification method (Sec.~\ref{ss:dmdt}) is to inform on when the next
observation will be most discriminatory for different classes; we need to have
separate means for obtaining observations. (3)~Depending on the nature of the
transient, different cadences are needed for follow-up (e.g.\ SNe need the
follow-up to be denser near the peak) and this can only be accomplished by
having access to telescopes with follow-up capabilities. (4)~Most crucially
though, since spectroscopic follow-up, the final arbiter, cannot be carried
out in every case, it is the early follow-up that can quickly determine if the
transient candidate is worthy of further observations (because it is an
outlier, or belongs to a rarer class) or it is one of the run-of-the-mill
types and can be safely put on a back-burner.

\begin{figure}
\centerline{\includegraphics[width=6cm]{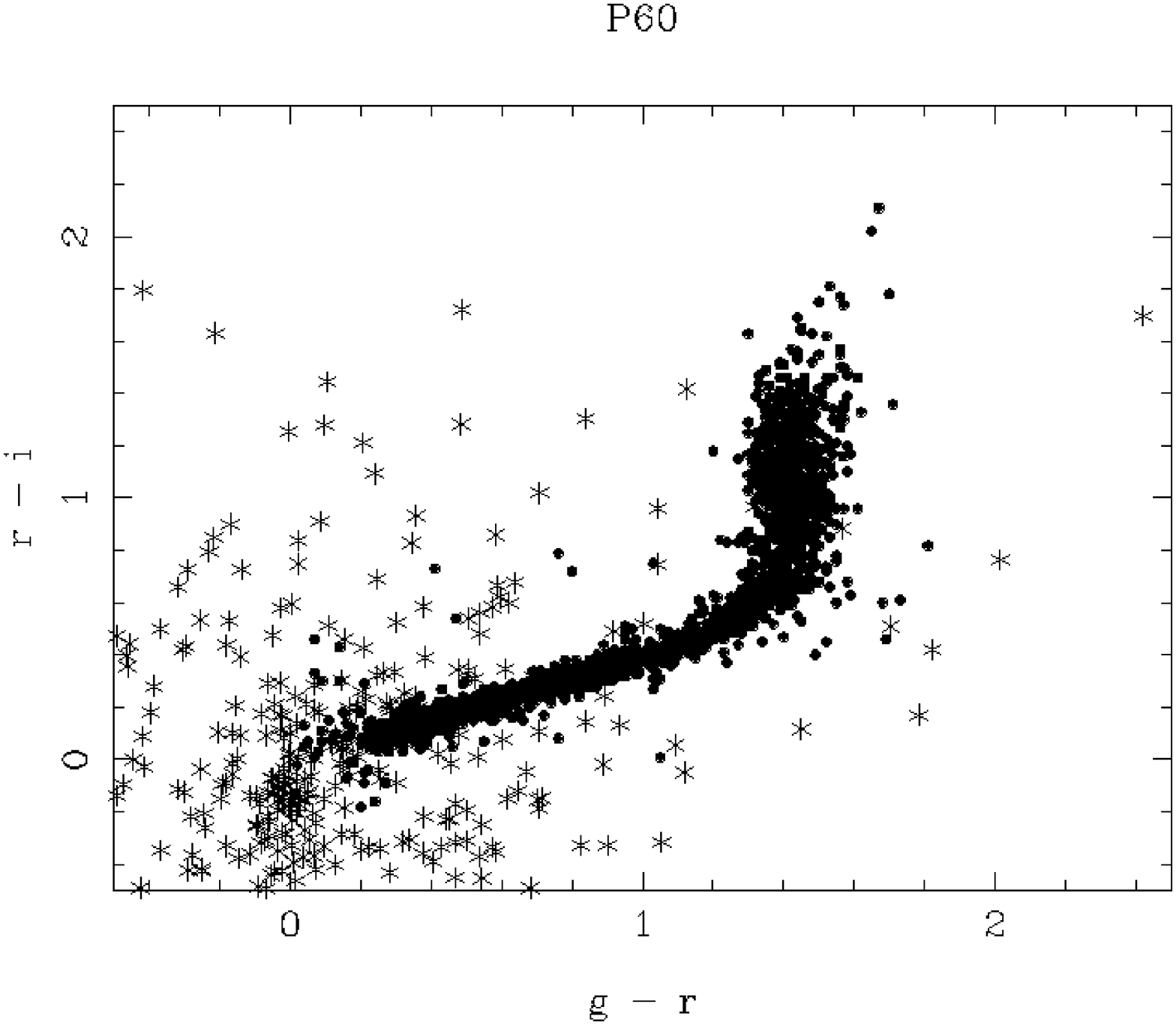} \qquad
\includegraphics[width=6cm]{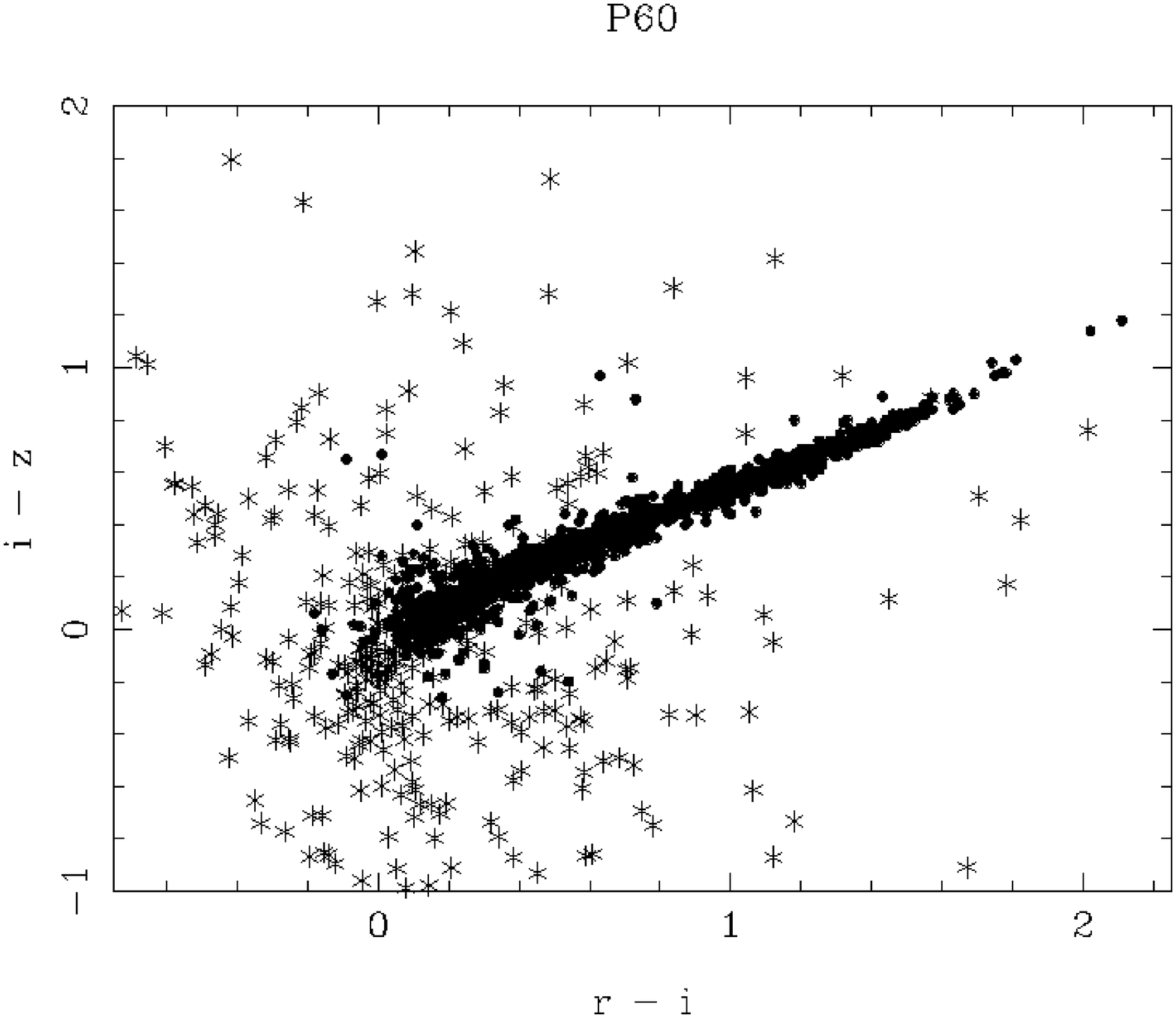} }
\caption{Distribution of colours from P60 follow-up. The locus is typical
non-variable stars. All epochs are plotted together. When different epochs for
a single object are plotted and connected as a function of time, one can see
the evolution of colours. As the data-set grows this provides vital information
to build priors for different classes.\label{f:p60}}
\end{figure}

With  all these in mind we have been carrying out follow-up from the 
P1.5m telescope in $g, r, i, z$ filters. This has allowed us to choose
objects for spectroscopic follow-up from telescopes such as the
IUCAA Girawali Observatory (IGO) 2-m, Palomar 5-m and Keck 10-m. It
has also contributed to various priors that form inputs to the Bayesian
Networks and provided sample LCs for the ($dm, dt$) method. Fig.~\ref{f:p60}
shows a stellar locus with colours from various transients from P1.5m
superimposed.

A variety of follow-up telescopes are needed (e.g.\ different apertures,
instruments, wavelength coverages etc.) for optimal follow-up of a range of
transients. We are working on another Bayesian tool that can provide the best
match for a given transient (based on whatever early parameters are available)
and one of several telescope+instrument pairs. For a given initial probability
distribution for different object types, the tool estimates best available
telescope and instrument combination that will disambiguate between the
different classes. In order to collect data for the network (besides the
reasons stated above) we have been obtaining follow-up epochs from IGO 2-m,
SMARTS 1.3-m, NMSU 1-m etc. We will soon have data from SAAO 1.9-m as well.

Gaia is slated to be launched in 2012. The magnitude distribution for the
transients found by Gaia is expected to be similar to that of CRTS. Keeping
that in mind a program is being initiated to observe CRTS transients with
various European telescopes in various states of automation. The open nature of
CRTS makes it ideal for such a test-bed. The network will be developed using
skyalert and VOEvents.

As needed, various other telescopes are invoked depending on the nature of the
transient (e.g.\ the Expanded Very Large Array (EVLA), HST and the Giant Metrewave
Radio Telescope (GMRT) were used for following CSS100217
described in Sec.~\ref{ss:nature}). For blazars follow-up observations are also
obtained from the 40-m OVRO radio telescope in the $15.0 \pm 1.5$ GHz band.

\subsection{Incorporating contextual information}\label{ss:context}

Contextual information can be highly relevant to resolving competing
interpretations: for example, the light curve and observed properties of a
transient might be consistent with it being a cataclysmic variable star, a
blazar, or a supernova. If it is subsequently known that there is a galaxy in
close proximity, the supernova interpretation becomes much more plausible. Such
information, however, can be characterized by high uncertainty and absence, and
by a rich structure: if there were two galaxies nearby instead of one then
details of galaxy type and structure and native stellar populations become
important, e.g.\ is this type of supernova more consistent with being in the
extended halo of a large spiral galaxy or in close proximity to a faint dwarf
galaxy?  The ability to incorporate such contextual information in a
quantifiable fashion is highly desirable. We have been compiling priors for
such information as well. These then get incorporated into the Bayesian network
(of Sec.~\ref{ss:BN}).

We are also investigating the use of crowdsourcing (`citizen science') as a
means of harvesting the human pattern recognition skills, especially in the
context of capturing the relevant contextual information, and turning them into
machine-processable algorithms. A methodology employing contextual knowledge
forms a natural extension to the logistic regression and classification methods
mentioned above. This is going to be necessary for larger future surveys when
we enter parameter spaces not explored before.

Ideally such knowledge can be expressed in a manipulable fashion within a sound
logical model, for example, it should be possible to state the rule that `a
supernova has a stellar progenitor and will be substantially brighter than it
by several orders of magnitude' with some metric of certainty and infer the
probabilities of observed data matching it. Markov Logic Networks (MLNs) are
such a probabilistic framework using declarative statements (in the form of
logical formulae) as atoms associated with real-valued weights expressing their
strength. The higher the weight, the greater the difference in log probability
between a world that satisfies the formula and one that does not, all other
thing being equal. In this way, it becomes possible to specify `soft' rules
that are likely to hold in the domain, but subject to exceptions -- contextual
relationships that are likely to hold such as supernovae may be  associated
with a nearby galaxy or objects closer to the Galactic plane may be stars. A
MLN defines a probability distribution over possible worlds with weights that
can be learned generatively or discriminatively: it is a model for the
conditional distribution of the set of query atoms $Y$ given the set of
evidence atoms $X$. Inferencing consists of finding the most probable state of
the world given some evidence or computing the probability that a formula holds
given a MLN and set of constants, and possibly other formulae as evidence. Thus
the likelihood of a transient being a supernova, depending on whether there was
a nearby galaxy, can be determined. The structure of a MLN --  the set of
formulae with their respective weights -- is also not static but can be revised
or extended with new formulae either learned from data or provided by third
parties. In this way, new information can easily be incorporated. Continuous
quantities, which form much of astronomical measurements, can also be easily
handled with a hybrid MLN.

These methods are in line with our philosophy that given the scale of the data
sets in near future there will not be enough humans to look at all possible
candidates and we will need programs that combine the brute force of computers
and the acumen of humans.

\subsection{Combining the classifiers}

A given classifier can not cater to all classes, nor to all types of inputs.
That is the primary reason why multiple types of classifiers have to be
employed in the complex task of classifying transients in real time. Presence
of different bits of information trigger different classifiers. In some cases
more than one classifier can be used for the same kinds of inputs. An essential
task, then, is to derive an optimal event classification, given inputs from a
diverse set of classifiers such as those described above. A fusion module is
used to accomplish this. However, the job of the fusion module viz. combining
different classifiers with different number of output classes and in presence
of error-bars is a non-trivial task and still being worked upon.
\label{ss:fusion}

\subsection{Citizen science}

We saw in Sec.~\ref{ss:context} how citizen science related to contextual
information is necessary for future surveys. We describe here another type of
citizen science, one involving regular monitoring of a large number of galaxies
for possible supernovae.

The main CRTS pipeline for transients is catalog-based. Transients can also be
found using the technique of image subtraction. This involves matching new
observations with either an older observation, or a deeper co-added image
(\citealt{Tom96}; \citealt{Dra99}). If the images are properly matched, transients
stand out as a positive residual. This is also useful when sources are blended
and is used in supernova searches and in crowded fields routinely
\citep{Ald02}. When used with white light, the difference images tend to have
bipolar residuals thus leading to false detections as well as missed
transients. We have been experimenting with these to look for supernovae in
galaxies using citizen science where a few amateur astronomers regularly look
at the galaxy images along with the residuals presented to them and by
answering a series of questions can determine if one of the candidates is
likely to be a genuine supernova. A few tens of supernovae have been found in
this fashion (see \citet{Pri11} for an example, and
\texttt{http://nesssi.cacr.caltech.edu/catalina/current.html} for a list).
Users are listed as official discoverers of any supernovae that they report,
provided that we can confirm that they are real, not already known, and they
have not previously been reported to us.

\subsection{CRTS transient event publishing}\label{ss:voevent}

To publish information on the transients in real time, CRTS uses VOEvents, an
international XML standard. A VOEvent \citep{Wil07} packet contains the basic
necessary information about the event like the time, location, magnitude, and
so on in sections marked ``who, what, where, when, how, why''  etc. These bits
are sufficient to initiate follow-up. The follow-up can be active, i.e.\ new
observations from a radio telescope or a spectrum, or it can be passive
e.g.\ querying an archival dataset for a lightcurve at that location or a
program that takes in whatever bits of information are available and returns a
verdict, say, the class of the object with associated probability values. The
information returned by each of these follow-ups get annotated to the main
entry. These annotators quote the id of the original event so that together
they form a cohesive portfolio for the transient.

The current follow-ups include observations from telescopes like the 
P1.5m, SMARTS 1.3-m, IGO 2-m, OVRO 40-m radio telescope (active) as well as
distances to and magnitudes of nearest star, galaxy, radio source etc. from a
variety of surveys; image cutouts from DPOSS, PQ, CRTS; past CRTS lightcurve;
basic classification; more informed classification based on some of the
follow-up information (passive).

Humans as well as computers and telescopes can subscribe to each of the CRTS
streams (CRTS for CSS, CRTS2 for MLS and CRTS3 for SSS). That way automated
follow-up can be done. In addition, one can set up arbitrarily complex filters
on these subscriptions so that one will get notified only under specific
circumstances. Some basic scenarios include (a) the CRTS stream produces a
transient with $g{-}r > 3$, or (b) there is a radio source within
$3''$, or (c) there is a galaxy brighter than 18th mag within
$10''$. This allows easy monitoring of specific classes of
objects. Different telescopes can thus be configured to receive only the
transients they are capable of following (based on, for example, mag, RA, Dec
limits.).

All the information is also available in the form of rich webpages, to which
expert comments can be added. One of the future plans includes running
semantic harvesting on the comments as well as entire portfolios to glean
higher level connections not captured in the basic annotators and to interface
with Virtual Observatory (VO) initiatives like VOSpace leading to a VO
Transient Facility. The list of transients and their portfolios can be found at
\texttt{http://www.skyalert.org/}.

%------------------------------------------------------------------------------%
\section{Concluding comments and future plans}

Surveys like CRTS already illustrate the great scientific richness and promise
of time domain astronomy, signaling even more exciting discoveries to come as
we move from the current terabyte regime to the petabyte regime of the near
future. The growing data rates require a strong cyber-infrastructure to match.
The time domain astronomy is an astronomy of telescope and computational/data
systems combined.

As we are moving ahead, there are several lessons learned worth emphasizing:
\begin{itemize}
\item The problem of a comprehensive follow-up of transient events is probably
the single greatest bottleneck at this time. Most of the science comes from
the follow-up observations, especially spectroscopy, and we are already
overwhelmed by the sheer numbers of the potentially interesting transients.
With CRTS, we estimate that only $\sim$10\% of the potentially interesting
events are followed up by anyone. This problem will grow by a several orders
of magnitude as we move into the LSST and SKA era.
\item The available follow-up assets (e.g.\ large enough telescopes for
spectroscopy) are unlikely to keep pace with the event discovery rates. Which
events, among the many, are worthy of the costly or resource-limited follow-up?
An essential enabling technology is thus the ability to automatically classify
and prioritize events, missing none of the interesting ones, and not saturating
the system with false alarms. This is a highly non-trivial problem, as
described above, and yet, it is the key for an effective, complete, and
responsible scientific exploitation of the synoptic sky surveys, both current
and forthcoming. A better community coordination of the follow-up efforts is
also important.
\end{itemize}

As for the CRTS survey itself, several ongoing and future developments may be
of interest:
\begin{itemize}
\item We are currently producing a database of about half a billion light
curves of all objects detected in multiple epochs over the entire survey area.
This will be an unprecedented resource for an archival exploration of the time
domain. We are starting to systematically characterize and analyze these light
curves. Also, as we have already demonstrated, archival light curves are
essential for the rapid characterization of newly discovered events.
\item Our co-added images reach fainter than $r \sim 23$ mag over most of the
survey area, i.e.\ $\sim 3/4$ of the entire sky. This will be another valuable
asset for the community.
\item The current CRTS transient detection threshold is set deliberately high,
in order to pick the most dramatic, high-contrast events; and even so, we can
follow-up only a small fraction of them. We plan to lower this threshold, thus
increasing the significant event discovery rate by an order of magnitude.
Combined with the archival light curves, this will also broaden the
astrophysical variety of objects and phenomena studied.
\item We are also in the process of cross-correlating CRTS sources with those
found at other wavelengths, e.g.\ in radio, or at high energies. This will
certainly produce a number of previously uncatalogued blazars and other AGN,
and possibly other types of objects as well (Mahabal et al., in preparation).
\end{itemize}

In summary, CRTS is a multi-faceted community asset for exploration of the time
domain. While the currently funded survey ends in late 2012, we hope that it
will be continued as an even more rewarding, larger effort.

%------------------------------------------------------------------------------%
\section*{Acknowledgements}

We wish to thank numerous collaborators who have contributed to the survey and
its scientific exploitation so far. CRTS is supported by the NSF grant
AST-0909182, and in part by the Ajax Foundation. The initial support was
provided by the NSF grant AST-0407448, and some of the software technology
development by the NASA grant 08-AISR08-0085. The analysis of the blazar data
was supported in part by the NASA grant 08-FERMI08-0025. Education and public
outreach activities are supported in part by the Microsoft Research WorldWide
Telescope team. The CSS survey is supported by the NASA grant NNG05GF22G. 
Some VOEvent related work was supported by NSF grant OCI-0915473.  We
are grateful to the staff of Palomar, Keck, and other pertinent observatories
for their expert help during our follow-up observations. Event publishing and
analysis benefits from the tools and services developed by the U.S. National
Virtual Observatory (now Virtual Astronomical Observatory).

%------------------------------------------------------------------------------%

\label{lastpage}
%------------------------------------------------------------------------------%
\end{document}